\documentclass[aps,secnumarabic,amsmath,amssymb,12pt]{revtex4}

\usepackage{bm}

\begin{document}

\title{Possible origin of Larson's lows}

\author{A.W.~Zaharow}
\affiliation{Ufa state petrotechnical university, Ufa, Russia}
\email{zaharow@gmail.com}

\date{April 10, 2007}

\numberwithin{equation}{section}

\begin{abstract}
It was found that approximately constant column densities of giant
molecular clouds (Larson's low) can be explained as cloud existence condition
in external (galactic) gravitational field. This condition can be also 
applied to objects (clumps and cores) embedded into the cloud and its
gravitational field. Derived existence condition do not rely on any internal
dynamic of a cloud and embedded objects.
\end{abstract}

\maketitle
\thispagestyle{empty}

\section{I\lowercase{ntroduction}}\label{S:intro} 
Giant molecular clouds (GMCs) play a crucial role in the star formation process
(\cite{McL9605},\cite{McK9901},\cite{McK9902},\cite{McK9903},\cite{Pad0701}).
GMCs are complex objects with masses $\sim 10^{5-6}M_{\odot}$,
diameters $\sim 50 pc$ and average densities $n_{H_2}\sim 10^2 cm^{-3}$
(e.g. \cite{Bl93}). GMCs are generally gravitationally bound and may contain
several sites of star formations. Equilibrium of self-gravitating gas was
theoretically investigated in many works (e.g. \cite{McK9903},\cite{Zah06} and references in
these papers). Internal structure of GMS is usually
very complicated. The inhomogeneous structure of cloud could be described as
a set of descrete clumps (\cite{Bl86}). These clumps themselves contain dense
cores with densities $n\sim 10^{3-5} cm^{-3}$. The one point of view at this
time is that this cloud structure is due to supersonic turbulence. Remarkably
enough, the properties of cloud complexes are rather simply interrelated.
Total masses, mean densities and average velosity dispersions vary with sizes
(effective radii) roughly as $M \varpropto R^2$,$\rho \varpropto R^{-1}$,
$\sigma \varpropto R^{1/2}$ (Larson's lows) (\cite{Lar81}).

In this paper I propose simple hypothesis to explain relation between
masses and sizes of clumps and cores, embedded into clouds.


\section{l\lowercase{arson's lows}}\label{S:ll}
Let us briefly consider some of the most salient characteristics of GMCs
summarized by Larson (\cite{Lar81}). See alsow (\cite{McK9901}). The first
relation is the line width-size relation: molecular clouds are supersonically
turbulent with line widths $\Delta v$ that increase as a power of size,
$\Delta v \varpropto R^p$. Larson himself estimated that $p \simeq 0.38$.
Subsequent work has distinguished between the relation valid for a collection
of GMCs and that valid within individual GMC. Within low-mass cores, Caselli
\& Myers (\cite{Cas95}) found that the nonthermal velocity dispersion is
\begin{equation} \label{eq:sig}
\sigma_{nt} \simeq 0.55R^{0.51}_{pc} \mathrm{~km ~s^{-1}}\, ,
\end{equation}
which is near to relation 
\begin{equation} \label{eq:sig1}
\sigma \varpropto R^{1/2} \,.
\end{equation}

Larson's second result was that GMCs and clumps within them are gravitationally
bound. It implies that 
\begin{equation} \label{eq:sig2}
\sigma^2 \propto GM/R \,.
\end{equation}
This relation is result
of virial equilibrium.

His third conclusion was that all GMCs have about the same column density
\begin{equation} \label{eq:dens}
N \simeq \mathrm{const} \,.
\end{equation}
 Column density is defined as 
\begin{equation} \label{eq:dens1}
N \propto M/R^2 \propto nR \,.
\end{equation}
As Larson pointed
out, only two of these conclusions are independent: any one of them can be
derived from other two.
Opposite to the second conclusion the first and the second Larson's lows
have no evident explanation. I think this explanation may be as follows.

Embedded clumps and cores moves through the cloud in its gravitational field
and are subject of turbulent motion and various accelerations. If their own 
gravitational field is not enough to hold on these objects they must
rapidly decay. It is easy to derive necessary condition for confinement gaseous body with
mass $M$ and radius $R$ in external gravitational field of cloud with mass
$M_{cl}$ and radius $R_{cl}$. The variation of self gravitational potential
$\Delta \phi$ must be greater then  that of the cloud on size $R$. Bearing in
mind that
\begin{equation} \label{eq:spot}
\Delta \phi \simeq GM/R 
\end{equation}
and
\begin{equation} \label{eq:cpot}
\Delta \phi_{cl} \simeq R \frac{d}{dR_{cl}}GM/R_{cl} \simeq GMR/R_{cl}^2
\end{equation}
we get
\begin{equation} \label{eq:pot1}
GM/R \ge GM_{cl}R/R^2_{cl}
\end{equation}
or equivalently
\begin{equation} \label{eq:pot2}
M/R^2 \ge M_{cl}/R^2_{cl} \,.
\end{equation}
Inequality (\ref{eq:pot2}) is the main result of this work. It is important for
undestanding the third Larson's low (\ref{eq:dens}). Inequality (\ref{eq:pot2}) 
is strong only for very massive and compact objects within the cloud which
are hard to generate in turbulent motion. So inequality can not be strong
for almost all objects. Further, in expression (\ref{eq:cpot}) we did not
take into account placement of object within the cloud. Summarizing all 
we can generalize third  Larson's low in the form
\begin{equation} \label{eq:mylr}
\frac{M}{R^2} \approx C \frac{M_{cl}(r)}{r^2} \,,
\end{equation}
where $r$ is the dinstance of object from the center of cloud, $M_{cl}(r)$
is the mass of cloud within $r$, $C$ is the non-dimensional constant of order
one. As far as clouds itself are concerned they move in galactic gravitational
field through the interstellar gaseous media (ISM) and we can get for them
analogous condition
\begin{equation} \label{eq:mylrcl}
\frac{M_{cl}}{R^2_{cl}} \approx C \frac{M_{gal}(r)}{r^2} \,,
\end{equation}
where $r$ is the dinstance of cloud from the center of galactic, $M_{gal}(r)$
is the mass of galactic within $r$.


\section{s\lowercase{ummary}}\label{S:sum}
We have seen that it is possible to understand the third Larson's low
and some properties of molecular clouds and objects within them in terms
of their existence conditions in external gravitational field. In particular,
we have seen possible reason why the clouds have approximately constant
column densities. I have to stress now that conditions (\ref{eq:mylr}) and 
(\ref{eq:mylrcl}) do not rely on any internal dynamic of cloud. It will be
also interesting  to examine relations (\ref{eq:mylr}) and (\ref{eq:mylrcl})
with astronomical data sets.

\end{document}